\documentclass{ws-ijmpd}

\begin{document}

\markboth{F. Saif}
{Coherent Acceleration of Material Wavepackets}

%
\catchline{matter waves}{}{}{}{}
%

\title{Coherent Acceleration of Material Wavepackets}

\author{Farhan Saif}

\address{Department of Electronics, Quaid-i-Azam University,
Islamabad 45320, Pakistan.\\
saif@fulbrightweb.org}

\author{Pierre Meystre}

\address{Department of Physics, The
University of Arizona, Tucson, AZ 85721, USA.}

\maketitle

\begin{history}
\received{Day Month Year}
\revised{Day Month Year}
\comby{Managing Editor}
\end{history}

\begin{abstract}
We study the quantum dynamics of a material wavepacket bouncing
off a modulated atomic mirror in the presence of a gravitational
field. We find the occurrence of coherent accelerated dynamics for
atoms. The acceleration takes place for certain initial phase space
data and within specific windows of modulation strengths. The
realization of the proposed acceleration scheme is within the
range of present day experimental possibilities.
\end{abstract}


\section{Introduction}

Accelerating particles using oscillating potentials is an area of extensive research\cite{saifpla}, first triggered by the ideas of Fermi on the origin of cosmic rays. In his seminal paper 'On the origin of cosmic rays', he stated that ``cosmic rays are originated and accelerated
primarily in the interstellar space of the galaxy by {\it
collisions} against moving magnetic fields''\cite{kn:fermi}.
This understanding lead to the development of two
major models: The Fermi-Ulam accelerator, which deals with the
bouncing of a particle off an oscillating surface in the presence
of another fixed surface parallel to it; and the
Fermi-Pustyl'nikov accelerator, where the particle bounces off an
oscillating surface in the presence of gravity. In the case of the
Fermi-Ulam accelerator~\cite{kn:lieb,kn:lilico} it was shown that
the energy of the particle remains bounded and the unlimited
acceleration proposed by Fermi is absent \cite{zaslavskii}. In the
Fermi-Pustyl'nikov accelerator, by contrast, there exists a set of
initial data within specific domains of phase space that result 
in trajectories speeding up to
infinity. 

In recent years the acceleration of laser-cooled atoms has become a topic of 
great interest for applications such as atom interferometry and  
the development of matter-wave based inertial sensors. 
Possible schemes of matter-wave acceleration have been proposed and studied. 
For example, a Bose Einstein condensate in a frequency-chirped optical lattice\cite{pott}
and an atom in an amplitude modulated optical lattice in 
the presence of a gravitational field 
display acceleration\cite{accelm}. The
$\delta$-kicked accelerator in the latter case operates for certain sets of initial
data that originate in stable islands of phase space.

Here, we discuss an experimentally realizable technique to accelerate a material wavepacket in a coherent fashion. It consists of an atom optics version of the
Fermi-Pustyl'nikov accelerator~\cite{Saif 1998} where a cloud of
ultracold atoms falling in a gravitational field bounces off a
spatially modulated atomic mirror. This scheme is different from
previous accelerator schemes in the following ways: {\it i}) The
regions of phase space that support acceleration are located in
the mixed phase space rather than in the islands of stability (or
nonlinear resonances); {\it ii}) The acceleration of the
wavepacket is coherent; {\it iii}) It occurs only for certain
windows of oscillation strengths.

\section{The Model}

We consider a cloud of laser-cooled atoms that move along the
vertical $\tilde z$-direction under the influence of gravity and
bounce back off an atomic mirror~\cite{kn:amin}. This mirror is
formed by a laser beam incident on a glass prism and undergoing
total internal reflection, thereby creating an optical evanescent
wave of intensity $I(\tilde z)=I_0\exp(- 2k \tilde z)$ and
characteristic decay length $k^{-1}$ outside of the prism.

The laser intensity is modulated by an acousto-optic modulator as
\cite{kn:sten} $I(\tilde z,\tilde t) = I_0\exp(- 2k\tilde z
+\epsilon\sin\omega\tilde t)$, where $\omega$ is the frequency and $\epsilon$ the amplitude of
modulation. The laser frequency is tuned far from any atomic
transition, so that there is no significant upper-state atomic
population. The excited atomic level(s) can then be adiabatically
eliminated, and the atoms behave for all practical purposes as
scalar particles of mass $m$ whose center-of-mass motion is
governed by the one-dimensional Hamiltonian
\begin{equation}
\tilde{H}=\frac{\tilde{p}^2}{2m}+mg\tilde{z}
+\frac{\hbar\Omega_{\rm eff}}{4} e^{-2k\tilde{z}+
\epsilon\sin\omega\tilde t}, \label{ham}
\end{equation}
where $\tilde p$ is the atomic momentum along $\tilde{z}$ and $g$
is the acceleration of gravity.

We proceed by introducing the dimensionless position and momentum
coordinates $z\equiv\tilde{z}\omega^2/g$ and
$p\equiv\tilde{p}\omega/(mg)$, the scaled time
$t\equiv\omega\tilde t$, the dimensionless intensity
$V_0\equiv\hbar\omega^2\Omega_{\rm eff}/(4mg^2)$, the
steepness $\kappa\equiv 2kg/\omega^2$, and the modulation
strength $\lambda\equiv\omega^2\epsilon/(2kg)$ of the evanescent
wave field.

When extended to an ensemble of non-interacting particles, the
classical dynamics obeys the condition of incompressibility of the
flow~\cite{kn:lieb}, and the phase space distribution
function $P(z,p,t)$ satisfies the Liouville equation.
In the absence of mirror modulation, the atomic dynamics is
integrable. For very weak modulations the incommensurate motion
almost follows the integrable evolution and remains rigorously
stable, as prescribed by the KAM theorem. As the modulation
increases, though, the classical system becomes chaotic.

In the quantum regime, the atomic evolution obeys the
corresponding Schr\"odinger equation.
The commutation relation,
$[z,p] ={\it i}(\omega^3 /mg^2) \hbar\equiv
{\it i}k^{\hspace{-2.1mm}-}$, naturally leads to the introduction of the dimensionless Planck constant, $k^{\hspace{-2.1mm}-}\equiv\hbar\omega^3/(mg^2)$. It can
easily be varied by changing for instance $\omega$, thereby permitting to study 
the transition from the 
semiclassical to the purely quantum dynamics of the atoms.

\section{Accelerated Dynamics}

The classical version of the is characterized by 
the existence of a set of initial conditions resulting in trajectories that accelerate without bound\cite{kn:pust}. More precisely, the classical evolution of the Fermi
accelerator displays the onset of global diffusion above a
critical modulation strength $\lambda_l=0.24$\cite{kn:lilico},
while the quantum evolution remains localized until a larger value
$\lambda_u$ of the modulation\cite{Saif 1998,kn:benv,kn:oliv}.
Above that point both the classical and the quantum dynamics are
diffusive. However, for specific sets of initial conditions that
lie within phase space disks of radius $\rho$, accelerating
modes appear for values of the modulation strength $\lambda$
within the windows~\cite{kn:pust}
\begin{equation}
s\pi \leq \lambda < \sqrt{1+(s\pi)^2},  \label{eq:lcon}
\end{equation}
where $s$ can take integer and half-integer values for the
sinusoidal modulation of the reflecting surface considered here.
\begin{figure}
\begin{center}
\includegraphics[scale=0.75]{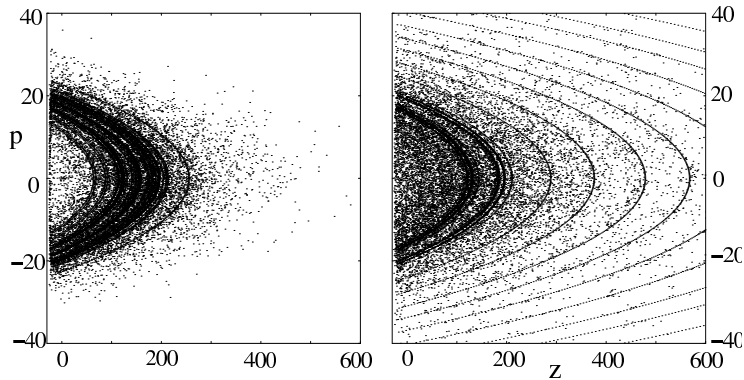}
\end{center}
\caption{Phase space evolution of a classical ensemble
of particles initially in a narrowly peaked Gaussian distribution
originating from the area of phase space that supports
accelerated trajectories. The initial distribution, centered at $\bar z=0$
and $\bar p=2\pi^2$ with $\Delta p(0) = \Delta z(0)=0.1$, is propagated for 
$\lambda=1$ (left) and $\lambda=1.7$ (right) for time $t=1000$.
The numerical calculations correspond to Cesium atoms of mass $m=2.2
\times 10^{-25}$Kg bouncing off an atomic mirror with an intensity
modulation of $\epsilon=0.55$. The modulation frequencies extend to the megahertz range, and $\kappa^{-1}=0.55$ $\mu$m. } \label{one}
\end{figure}

We found numerically that for a modulation 
strength outside the windows of Eq.~(\ref{eq:lcon}) the dynamics is dominantly diffusive. However,
as the fundamental requirement for the acceleration is met by chosing a 
modulation strength within the windows the 
ensemble displays a nondispersive and coherent acceleration, Fig.~\ref{one}. A small diffusive background results from a small part of the initial distribution which is residing outside the area of phase space supporting acceleration. 
This coherent acceleration restricts the momentum 
space variance $\Delta p$ which then remains very small
indicating the absence of diffusive dynamics\cite{SaifMeystre}. 

In the quantum case the Heisenberg Uncertainty Principle imposes a
limit on the smallest size of the initial wavepacket. Thus in
order to form an initial wavepacket that resides entirely within
regions of phase space leading to coherent dispersionless
acceleration, an appropriate value of the effective Planck constant
must be chosen, for example, by controlling the frequency
$\omega$\cite{tobe}. For a broad wavepacket, the coherent acceleration manifests itself
as regular spikes in the marginal probability distributions $P(p,t)= \int dx P(x,p,t)$
and $P(x,t)= \int dp P(x,p,t)$.

\begin{figure}[t]
\begin{center}
\includegraphics[scale=0.4]{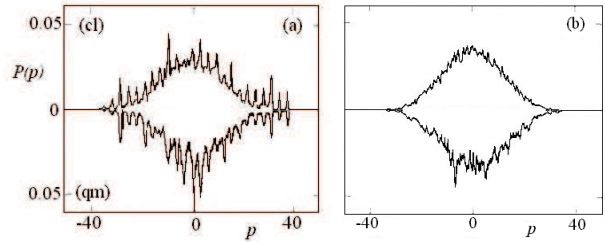}
\end{center}
\caption{Mirror images of the classical and quantum mechanical momentum distributions,
$P(p)$, plotted for (a) $\lambda=1.7$ within acceleration window, and (b) $\lambda=2.4$ outside the acceleration window. The
spikes in the momentum distribution for $\lambda=1.7$ are a
signature of coherent accelerated dynamics. The initial width of the momentum distribution is $\Delta p=0.5$ and the probability distributions are recorded after a scaled propagation time $t=500$. The initial probability distributions has variance $\Delta p=0.5$, which fulfills the minimum uncertainty relation. } \label{fig:two}
\end{figure}

This is illustrated in Fig.~\ref{fig:two} which shows the marginal probability distribution $P(p,t)$ for (a) $\lambda=1.7$ and (b) $\lambda=2.4$,
both in the classical and the quantum domains. In this example, the initial area of the
particle phase-space distribution is taken to be large compared to the
size of the phase-space regions leading to purely unbounded
dispersionless acceleration. The sharp spikes in, $P(p,t)$ appear when the modulation strength satisfies the condition of Eq.~(\ref{eq:lcon}), and gradually disappear as
it exits these windows. These spikes are therefore a
signature of the coherent accelerated dynamics. In contrast, the portions of the initial probability distribution originating from the regions of the phase space
that do not support accelerated dynamics undergo diffusive dynamics. 

From the numerical results of Fig.~\ref{fig:two},
we conjecture that the spikes are well
described by a sequence of gaussian distributions separated by a
distance $\pi$, both in momentum space and coordinate space. 
We can therefore express the complete time-evolved
wavepacket composed of a series of sharply-peaked gaussian
distributions superposed to a broad background due to diffusive
dynamics, such that
\begin{eqnarray}
P(p)&=&{\cal N} e^{-p^2/4\Delta p^2}\sum_{n=-\infty}^{\infty}
e^{-(p-n\pi)^2/4\epsilon^2},
\end{eqnarray}
where $\epsilon << \Delta p$, and $\cal N$ is a normalization constant.

\begin{figure}[t]
\begin{center}
\includegraphics[scale=0.55]{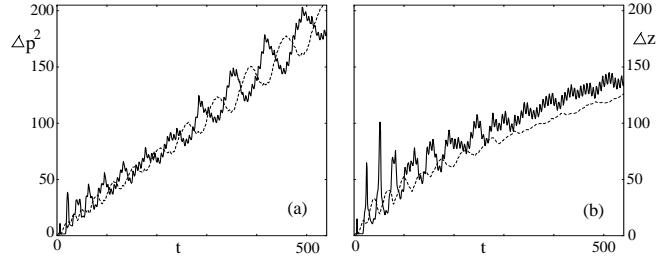}
\end{center}
\caption{(a) Square of the momentum variance (dark lines) and coordinate space 
variance (gray lines) as a function of time for $\lambda=1.7$. The coherent acceleration results in a breathing of the atomic wave packet, as evidenced by the out-of-phase oscillations of the variances.
(b) Dynamics for $\lambda=2.4$, a modulation strength that does not result in coherent acceleration. Note the absence of breathing in that case. Same parameters as in Fig.~\ref{fig:two}.} \label{fig:three}
\end{figure}

Further insight in the quantum acceleration of the atomic wavepacket is obtained by studying its temporal evolution. We find that within the window of acceleration the atomic wave packet displays a linear growth in the square of the momentum variance and in the coordinate space variance. 
Figure~\ref{fig:three} illustrates that for modulation strengths within the acceleration window, the growth in square of the momentum variance 
displays oscillations of increasing periodicity whereas the variance in
coordinate space follows with a phase difference of $180^o$. 
The out-of-phase oscillatory evolutions of $\Delta p^2$ and $\Delta z$ indicate a breathing of the
wavepacket and is a signature of the coherence in accelerated dynamics. As a final point we note that outside of the acceleration window the linear growth in the square of the momentum variance, a consequence of normal diffusion, translates into a $t^{\alpha}$ law, with $\alpha < 1$ which is a consequence of anomalous diffusion. 

\section{Summary}

We have investigated the classical and quantum evolution of atoms in a Fermi accelerator beyond the regime of dynamical localization where diffusive behavior occurs both in the classical and the quantum domains. We have identified the conditions leading to the
coherent acceleration of the atoms, and found signatures of the behavior both for an ensemble of classical particles and for a
quantum wavepacket. A quantum wavepacket with a broad initial variance, restricted by the Heisenberg Uncertainty Principle results in the coherent
acceleration occurring on top of a diffusive background. 

\section*{Acknowledgment}

This work is supported in part by the US Office of Naval Research,
the National Science Foundation, the US Army Research
Office, the Joint Services Optics Program, the National
Aeronautics and Space Administration, and J. William Fulbright foundation.

\end{document}